\documentclass[a4paper,10pt]{article}
\pdfoutput=1  

\usepackage{jheppub}  
\usepackage{amsmath}
\usepackage[T1]{fontenc}  
\usepackage{graphicx}
\usepackage{epstopdf}
\usepackage{bm}
\usepackage{epsfig}
\usepackage{graphics}
\usepackage{hyperref}
\usepackage{soul}

 \usepackage{ulem}

\newcommand{\Tr}{\text{Tr}}
\newcommand{\Li}{\text{Li}}
\newcommand{\ot}{\leftarrow}
\newcommand{\nn}{\nonumber}

\renewcommand{\vec}[1]{\bm{#1}}

\newcommand{\specialcellcenter}[2][c]{\begin{tabular}[#1]{@{}c@{}}#2\end{tabular}}

\def\({\left(}
\def\[{\left[}
\def\){\right)}
\def\]{\right]}

\title{Linearly polarized gluons at next-to-next-to leading order and the Higgs transverse momentum distribution}

\author[a]{Daniel Gutierrez-Reyes,}
\author[a,b]{Sergio Leal-Gomez,}
\author[a]{Ignazio Scimemi,}
\author[c]{Alexey Vladimirov}

\affiliation[a]{Departamento de F\'isica Te\'orica and IPARCOS,Universidad Complutense de Madrid (UCM),28040 Madrid, Spain}
\affiliation[b]{University of Vienna, Faculty of Physics, Boltzmanngasse 5, A-1090 Wien, Austria}
\affiliation[c]{Institut f\"ur Theoretische Physik, Universit\"at Regensburg,D-93040 Regensburg, Germany}   

\emailAdd{dangut01@ucm.es}
\emailAdd{sergiol95@univie.ac.at}
\emailAdd{ignazios@ucm.es}
\emailAdd{alexey.vladimirov@ur.de}
\preprint{UWThPh-2019-23}
\abstract{
We calculate the small-$b$ (or large-$q_T$) matching of transverse momentum dependent (TMD) distribution for linearly polarized gluons to the integrated gluon distributions at the next-to-next-to-leading order (NNLO). This is the last missing part for the complete NNLO prediction of the Higgs spectrum within TMD factorization. We discuss the numerical impact of the correction so derived to the $q_T$-differential cross-section for Higgs boson production and to the positivity bound for linearly polarized gluon transverse momentum distribution.}
 
\begin{document}
\maketitle

\section{Introduction}

The gluon-gluon fusion is the leading channel for the Higgs boson production in  hadron-hadron collisions \cite{Ellis:1975ap,Spira:1995rr,Djouadi:2005gi}. The transverse momentum dependent (TMD) factorization of Higgs production  has been  demonstrated to follow the same pattern as  the Drell-Yan/vector boson case in different frameworks \cite{Catani:1988vd,Collins:2011zzd,GarciaEchevarria:2011rb,Echevarria:2014rua,Vladimirov:2017ksc} and in this sense it has been reviewed in \cite{Echevarria:2015uaa}. Within the TMD factorization theorem, which  describes the Higgs production at small transverse momentum, there are two  dominant terms in the factorized cross-section. Those terms correspond to the fusion of unpolarized and the linearly polarized gluons \cite{Catani:2010pd,Boer:2011kf,Becher:2012yn}. Schematically, it reads
\begin{align}
\label{eq:start}
\frac{d\sigma}{d y d^2 \vec q_T}&=\frac{\sigma_{gg\to H}}{(2\pi)^2} 
 \int d^2 \vec b \;
e^{-i (\vec b \vec q_T)}
\Big(
f_{1,g} (x_A,\vec  b) f_{1,g} (x_B,\vec b) +
h^{\perp }_{1,g} (x_A,\vec b) h^{\perp }_{1,g} (x_B,\vec b)\Big)
\ ,
\end{align}
where $\sigma_{gg\to H}$ is the factorized gluon-gluon-Higgs cross-section, $x_{A,B}$ are the collinear fractions of gluon momenta, $f_1$ is the unpolarized gluon traverse momentum dependent parton distribution function (TMDPDF) and $h_{1,g}^\perp$ is the linearly polarized gluon TMDPDF (lpTMDPDF) that  was proposed as an independent distribution a long ago by Mulders and Rodrigues~\cite{Mulders:2000sh}.

In TMD factorization  each TMD distribution ($f_{1,g}$ and $h_{1,g}^\perp$ in this case) is an independent fundamental non-perturbative function. 
In order to sensibly construct a TMD for any practical  purpose it is  fundamental to include the asymptotical  small-$b$ limit, where each TMD distribution match to collinear parton distributions and the matching coefficient is calculable in  QCD perturbation theory \cite{Collins:2011zzd,GarciaEchevarria:2011rb,Becher:2010tm}. 
The modern state-of-the-art of perturbative calculations these matchings is the next-to-next-to-leading order (NNLO) of perturbative series, see \cite{Gehrmann:2014yya,Echevarria:2016scs,Gutierrez-Reyes:2017glx,Luo:2019hmp}. Such a high order is required because of the sizes of theoretical and experimental uncertainties, see e.g., \cite{Scimemi:2018xaf,Cruz-Martinez:2018rod}. Also, it is required for the use of NNLO TMD evolution, which is necessary to perform an accurate global analysis of high- and low-energy data \cite{Scimemi:2017etj,Bertone:2019nxa}. The small-$b$ limit of the unpolarized gluon TMDPDF, $f_1$, has been calculated at NNLO in \cite{Gehrmann:2014yya,Echevarria:2016scs}. However the small-$b$ limit of the lpTMDPDF,  $h_{1,g}^{\perp}$ is known only at one-loop  \cite{Becher:2012yn,Echevarria:2015uaa,Gutierrez-Reyes:2017glx} and as such it has been used  in ref.~\cite{Chen:2018pzu}\footnote{In ref.~\cite{Chen:2018pzu} the authors use the  differential cross section for Higgs production at NNLO which includes only the NLO matching coefficient for the linearly polarized gluons. This counting is different from the one required by  TMD factorization as explained in the text.}.

In this work, we fill this gap, providing the calculation of $h_{1,g}^{\perp}$ at two loops and estimating the impact of this correction on the Higgs transverse momentum spectrum. The calculation can be performed using the same techniques as in ref.~\cite{Echevarria:2015usa,Echevarria:2015byo,Echevarria:2016scs,Gutierrez-Reyes:2018iod}.

The necessity of  the present calculation comes from the fact that the perturbative counting in TMD formalism is slightly different from the one used in resummation approach.
In fact, using standard resummation, 
 see e.g. \cite{Bozzi:2005wk,Mantry:2009qz,deFlorian:2011xf,Becher:2012yn,Bizon:2018foh,Cruz-Martinez:2018rod},   the small-$b$ expansion is incorporated into the factorization formula, ignoring the non-perturbative TMD effects and one worries only about the perturbative expansion of the cross section. 
 The TMD factorization includes  the resummation for large enough $q_T$, however one has different requirements in the realization of the perturbative series.
  So, while in the usual resummation  the whole bracketed factor in eq.~(\ref{eq:start}) should be given at a certain perturbative order, in  TMD factorization  each distribution should be matched independently to its collinear counterpart at the same given order.  Both approaches are consistent with computing the small-$b$ expansion at the same order.  The case of linearly polarized gluon contribution to eq.~(\ref{eq:start}) is special because the tree-level matching  accidentally vanishes.  The counting of perturbative orders in TMD factorization is reported later in the text.

The result obtained in this work is relevant for many cases beyond the Higgs boson production. In particular, there are processes that  are also sensitive to lpTMDPDF and that are addressed in the literature \cite{Boer:2010zf,Metz:2011wb,Dominguez:2011br,Pisano:2013cya,Dumitru:2015gaa}. Among these it is worth a special mentioning the case of heavy-quark production \cite{Mukherjee:2016cjw,Boer:2017xpy,Efremov:2017iwh,Efremov:2018myn,Kishore:2018ugo,Echevarria:2019ynx,Marquet:2017xwy}, which is relevant at LHC, future Electron-Ion Collider (EIC) or the LHeC. Another important topic is the positivity bound for gluon TMDPDF derived in \cite{Mulders:2000sh},
\begin{align}
\label{eq:positivitybound}
|h_{1,g\leftarrow h}^\perp(x, \vec{q}_T)|/|f_{1,g\leftarrow h}(x,\vec{q}_T)|\leq 1.
\end{align}
This positivity bound is expected to saturate at small-$x$ due to the McLerran-Venugopalan model \cite{McLerran:1993ka}. Our calculation shows that this bound is easily violated by loop corrections but could be restored by non-perturbative corrections. In this way, the relation in eq.~(\ref{eq:positivitybound}) could be considered as a strong restriction on transverse momentum dependence of partons.
 
The two-loop calculation presented here is structured in a way similar to the case of unpolarized gluons, evaluated in \cite{Echevarria:2016scs}. We find it sufficient to recall the basic principles and notation in sec.~\ref{sec:Notation}, which can be skipped by the reader already acquainted with topical works. The computation has requested the calculations of several new master integrals which are reported in the appendix. The final result for the NNLO matching of $h_{1,g}^{\perp}$  onto collinear gluon PDF is presented in sec.~\ref{sec:rrrr}. The NNLO matching calculated here has been incorporated into \texttt{artemide} \cite{web}, which was used to perform a qualitative numerical estimation of lpTMDPDF to Higgs-production cross-section at  NNLO-N$^3$LL. The results of the phenomenological analysis are discussed in sec.~\ref{sec:Results}.

%%%%%%%%%%%%%%%%%%%%%%%%%%%%%%%%%%
\section{Gluon TMD distributions}
\label{sec:Notation}
%%%%%%%%%%%%%%%%%%%%%%%%%%%%%%%%%%%%%

\subsection{Definition}

The TMD distribution of gluons in a hadron is given by the following matrix element
\begin{align}
\label{def:TMD_OP_G}
\Phi_{g\leftarrow h, \mu\nu}(x,\vec b)&=\frac{1}{xp^+}\int \frac{d\lambda}{2\pi}e^{-ixp^+\lambda}
\\\nn \times &\langle P,S|
\bar T \left\{F_{+\mu}\(\lambda n+\vec b\)\tilde W_{n}\(\lambda n+\vec b\)\right\} 
T\left\{ \tilde W_{n}^{\dagger}(0)F_{+\nu}(0)\right\}|P,S\rangle,
\end{align}
where $n$ is a light-like vector, $F^{\mu\nu}$ is the gluon field strength tensor, and $\tilde W$ denotes the half-infinite Wilson line in the direction $n$
\begin{eqnarray}\label{def:Wline}
\tilde W_n(z)=P\exp\Big(ig\int_{-\infty}^0 d\sigma A_+(n\sigma+z)\Big).
\end{eqnarray}
The Wilson lines $\tilde W_n$ are taken in the adjoint representation of the gauge group.  We use the standard notation for the light-cone components of vector $v^\mu=n^\mu v^-+\bar n^\mu v^++g_T^{\mu\nu}v_\nu$ (with $n^2=\bar n^2=0$, $n\cdot\bar n=1$, and $g_T^{\mu\nu}=g_{\mu\nu}-n^\mu \bar n^\nu-\bar n^\mu n^\nu$).

The decomposition of the gluon TMD distribution over independent Lorenz structures contains 8 components \cite{Mulders:2000sh,Echevarria:2015uaa}. Two of these structures survive in the case of unpolarized hadron
\begin{align}
\label{TMD_G_dec}
&\Phi_{g\ot h}^{\mu\nu}(x,\vec b)=
-\frac{g_T^{\mu\nu}}{2(1-\epsilon)}f_{1,g\leftarrow h} (x,\vec b)+h_{1,g\leftarrow h}^{\perp} (x,\vec b)\(\frac{g_T^{\mu\nu}}{2(1-\epsilon)}+\frac{ b^\mu  b^\nu}{\vec b^2}\),
\end{align}
where $\vec b^2=-b^2>0$. For future necessity, the decomposition in eq.~(\ref{TMD_G_dec}) is given in $d=4-2\epsilon$-dimensions as it was defined in \cite{Gehrmann:2014yya,Gutierrez-Reyes:2017glx}. 
Both  $f_1$  and  $h_1^\perp$ contribute to the gluon-induced TMD processes on equal foot. Although these functions share some common properties, they are completely independent non-perturbative functions that are to be extracted from the experiment.

The usage of a $d$-dimensional definition for the decomposition in eq.~(\ref{TMD_G_dec}) is important for the following two-loop calculation because the $\epsilon$-dependent parts influence the result.
 The definition in eq.~(\ref{TMD_G_dec}) is the standard one
 %\footnote{We note however that in ref.~\cite{Gehrmann:2014yya} the  lpTMDPDF is defined with opposite sign. Here, we use the sign convention that is usually used, see e.g.~\cite{Mulders:2000sh,Becher:2012yn,Echevarria:2015uaa,Gutierrez-Reyes:2017glx}.} 
 \cite{Gehrmann:2014yya,Gutierrez-Reyes:2017glx} written such that the unpolarized part coincides with the standard definition of the unpolarized TMDPDF, see e.g. \cite{Collins:2011zzd,Echevarria:2016scs,Gehrmann:2014yya} (here dots denote the staple gauge link, as in (\ref{def:TMD_OP_G})),
\begin{eqnarray}
f_{1,g\ot h}(x,\vec b)&=&-g_T^{\mu\nu}\Phi_{g\leftarrow h, \mu\nu}(x,\vec b)
= \frac{1}{xp^+}\int \frac{d\lambda}{2\pi}e^{-ixp^+\lambda}
\langle P|F_{+\mu}\(\lambda n+\vec b\) ... F_{+\mu}(0)|P\rangle,
\end{eqnarray} 
whereas the linearly-polarized tensor is orthogonal to it. In turn the lpTMDPDF is given by
\begin{eqnarray}\label{eq:l-proj}
h^\perp_{1,g\ot h}(x,\vec b)&=&\frac{1}{1-2\epsilon}\(g_{T}^{\mu\nu}+2(1-\epsilon)\frac{b^\mu b^\nu}{\vec b^2}\)\Phi_{g\leftarrow h, \mu\nu}(x,\vec b).
\end{eqnarray}

Sometimes, one would like to use TMD distributions defined in the momentum space. The relation between coordinate and momentum representation is the usual one~\cite{Mulders:2000sh,Echevarria:2015uaa} (here in $d=4$ dimensions),
\begin{eqnarray}
\Phi_{g\leftarrow h, \mu\nu}(x,\vec k)&=&\int \frac{d^2 \vec b}{(2\pi)^2}e^{i(\vec b \vec k)}\Phi_{g\leftarrow h, \mu\nu}(x,\vec b)
\\\nn&
=&-\frac{g_T^{\mu\nu}}{2}f_{1,g\leftarrow h} (x,\vec k)+h_{1,g\leftarrow h}^{\perp} (x,\vec k)\(\frac{g_T^{\mu\nu}}{2}+\frac{k^\mu  k^\nu}{\vec k^2}\),
\end{eqnarray}
where
\begin{eqnarray}
f_{1,g\leftarrow h} (x,\vec k)=\int_0^\infty \frac{|\vec b|d|\vec b|}{2\pi}J_0(|\vec b||\vec k|)\,f_{1,g\leftarrow h} (x,\vec b),
\\
h^\perp_{1,g\leftarrow h} (x,\vec k)=-\int_0^\infty \frac{|\vec b|d|\vec b|}{2\pi}J_2(|\vec b||\vec k|)\,h^\perp_{1,g\leftarrow h} (x,\vec b).
\end{eqnarray}

\subsection{OPE at small-$b$}

At small-$b$ the TMD operator can be matched to the collinear operators by means of  operator product expansion (OPE). This relation is important because it constrains the  model for TMD distributions at small values of $b$. Moreover, at large values of $Q$, where the TMD evolution factor significantly suppress the large-$b$ part of the Fourier integral, the small-$b$ OPE provides the dominating input to the cross-section (see e.g.\cite{Becher:2012yn,Echevarria:2015uaa,Chen:2018pzu} for studies related to Higgs boson processes).

The systematic description of the small-$b$ OPE applied to TMD operators can be found in ref.~\cite{Scimemi:2019gge}. In the present case, it results into the following expressions
\begin{eqnarray}\label{OPE_unG}
f_{1,g\ot h}(x,\vec b;\mu,\zeta)&=&\sum_f \int_x^1 \frac{dy}{y} C_{g\ot f}(y,\vec b;\mu,\zeta;\tilde \mu)\,f_{1,f\ot h}\(\frac{x}{y},\tilde \mu\)+\mathcal{O}(\vec b^2)
\\\label{OPE_lpG}
h^\perp_{1,g\ot h}(x,\vec b;\mu,\zeta)&=&\sum_f \int_x^1 \frac{dy}{y} \delta^L\!C_{g\ot f}(y,\vec b;\mu,\zeta;\tilde \mu)\,f_{1,f\ot h}\(\frac{x}{y},\tilde \mu\)+\mathcal{O}(\vec b^2),
\end{eqnarray}
where the sum runs over the active parton flavors (quarks and gluon), and $f_1(x,\mu)$ is unpolarized collinear distributions defined as usual
\begin{eqnarray}\label{def:f1-PDF}
f_{1,q\ot h}(x,\mu)&=&\int \frac{d\lambda}{2\pi}e^{-ixp^+\lambda}
\langle P|\bar T\{\bar q\(\lambda n\) \tilde W_n(\lambda n)\}\frac{\gamma^+}{2}T\{\tilde W_n^\dagger(0)q(0)\}|P\rangle,
\\
f_{1,g\ot h}(x,\mu)&=&
\frac{1}{xp^+}\int \frac{d\lambda}{2\pi}e^{-ixp^+\lambda}
\langle P|\bar T \left\{F_{+\mu}(\lambda n)\tilde W_{n}(\lambda n)\right\} 
T\left\{ \tilde W_{n}^{\dagger}(0)F_{+\mu}(0)\right\}|P\rangle.
\end{eqnarray}
Concerning the notation, here and in the following we distinguish the unpolarized TMDPDF $f_1(x,\vec b)$ and unpolarized collinear PDF $f_1(x)$ by the number of arguments. The scales $\mu$ and $\zeta$ in eq.~~(\ref{OPE_unG},~\ref{OPE_lpG}) are the scales of TMD evolution discussed in the next section. The scale $\tilde \mu$ is the scale of OPE, that is not related to the TMD evolution scales and whose dependence  cancels in the convolution of coefficient function and collinear distribution.

The coefficient functions (also known as matching coefficients \cite{Collins:2011zzd}),  $C$ and $\delta^L\!C$, are to be calculated in QCD perturbation theory. The three-order calculation yields
\begin{eqnarray}
C_{g\ot f}(x,\vec b;\mu,\zeta;\tilde \mu)&=&\delta_{gf}\delta(1-x)+\mathcal{O}(a_s),
\\
\delta^L\!C_{g\ot f}(x,\vec b;\mu,\zeta;\tilde \mu)&=&\mathcal{O}(a_s),
\end{eqnarray}
where $a_s=g^2/(4\pi)^2$ is QCD coupling constant. Nowadays, the coefficients $C_{f\ot h}(x,\vec b)$ are known at $a_s^2$-order (NNLO) \cite{Gehrmann:2012ze,Gehrmann:2014yya,Echevarria:2015usa,Echevarria:2016scs}, whereas coefficients $\delta^L\!C_{f\ot h}(x,\vec b)$ are known at $a_s$-order (NLO\footnote{In literature related to TMD calculations, e.g. in refs.\cite{Echevarria:2015uaa,Gutierrez-Reyes:2017glx}, the orders of $\delta^L\!C_{f\ot h}$ are traditionally counted alike the unpolarized case. So, the linear $a_s$-terms  are denoted as NLO. Here we use the same convention.}) \cite{Becher:2012yn,Echevarria:2015uaa,Gutierrez-Reyes:2017glx}. In the following section we present NNLO expression for $\delta^L\!C_{g\ot f}$, which allows to consider these distributions at the same level of accuracy.

The  corrections to the OPE at higher powers of $b$ are unknown but at  large value of $\vec b^2$ the OPE becomes divergent. Thus, in practice, for the description of the TMD distributions one typically uses a phenomenological ansatz that matches the OPE results at small-$b$ to a non-perturbative input at large-$b$. It can be written in the form
\begin{eqnarray}\label{h_ansatz}
h^\perp_{1,g\ot h}(x,\vec b)&=&\sum_f \int_x^1 \frac{dy}{y} \delta^L\!C_{g\ot f}(y,\vec b)\,f_{1,f\ot h}\(\frac{x}{y}\)h^\perp_{1\text{NP}}(x,y,\vec b^2),
\end{eqnarray}
and a similar expression can be used for $f_1(x,\vec b)$ with a different $f_{1\text{NP}}(x,y,\vec b^2)$ and the corresponding matching coefficient. In eq.~(\ref{h_ansatz}) we omit scale variables, and the function $h^\perp_{1\text{NP}}$ is an arbitrary function with the only constraint
\begin{eqnarray}
\label{eq:lig}
\lim_{\vec b^2\to 0}h^\perp_{1\text{NP}}(x,y,\vec b^2)\simeq 1+\mathcal{O}(\vec b^2),
\end{eqnarray}
which is necessary to be consistent with the small-$b$ limit of the TMD.
A similar ansatz has been used also for the quark TMD, and the respective non-perturbative correction has been called $f_{NP}$ in  ref.~\cite{Scimemi:2017etj,Bertone:2019nxa}.
Up to now the non-perturbative correction to the quark TMD is the only one which has been extracted from data. In order to have some  phenomenological result here we also choose 
$f_{NP}=f_{1NP}=h^\perp_{1NP}$. We comment about the consistency of this choice in sec.~\ref{sec:Results}.

\subsection{Renormalization of TMDPDF}
\label{sec:soft}

The TMD operator contains ultraviolet (UV) and rapidity divergences. Both these divergences can be renormalized (the all-order proof of renormalization for rapidity divergences is given in ref.~\cite{Vladimirov:2017ksc}) by the corresponding renormalization factors. Hence, the renormalized (or physical) TMD distribution depends on two scales $\mu$ (the UV renormalization scale) and $\zeta$ (the rapidity divergences renormalization scale). The renormalized expression for the TMD distribution $\Phi_{g\ot h}$ reads
\begin{eqnarray}\label{TMD_ren}
\Phi_{g\leftarrow h}(x,\vec b;\mu,\zeta)
=Z^{\text{TMD}}_g(\mu,\zeta|\epsilon)R_g\(\vec b,\mu,\zeta|\epsilon,\frac{\delta^+}{p^+}\)\Phi^{\text{unsub.}}_{g\leftarrow h}\(x,\vec b|\epsilon,\frac{\delta^+}{p^+}\),
\end{eqnarray}
where $\Phi^{\mu\nu;\text{unsub.}}_{g\leftarrow h}$ denotes the bare or unsubtracted TMD distribution, either $f_1$ either $h_1^\perp$, since the TMD renormalization is independent of polarization properties. In eq.~(\ref{TMD_ren}) we present explicitly the dependence on regularization parameters: $\epsilon$ is the parameter of dimensional regularization ($d=4-2\epsilon$) that regularizes UV divergences,  that are renormalized by the factor $Z_g$; $\delta$ is the parameter of $\delta$-regularization \cite{Echevarria:2015byo,Echevarria:2016scs}  which regularizes rapidity divergences  that are renormalized by the factor $R_g$. The renormalization factors  $Z_g$ and $R_g$ are ordered such that the renormalization of rapidity divergences is made before to the renormalization of UV divergences as it was done in similar NNLO calculations \cite{Echevarria:2015usa,Echevarria:2016scs,Gutierrez-Reyes:2018iod}. The final result is independent of the subtraction order.

The rapidity renormalization factor can be related to the TMD soft factor \cite{Vladimirov:2017ksc}, which is the vacuum expectation value of certain Wilson loop \cite{GarciaEchevarria:2011rb,Collins:2011zzd,Vladimirov:2017ksc},
\begin{align} \label{eq:softf}
S(\vec b) &=
\frac{{\Tr}_\text{color}}{N_c}
\langle 0|\[W_n^{T\dagger} \tilde W_{\bar n}^T \](\vec b)
\[\tilde W^{T\dagger}_{\bar n} W_n^T\](0)|0\rangle,
\end{align}
where $\tilde W_n$ and ${\tilde W}_{\bar n}$ are Wilson lines along $n$ and $\bar n$ (\ref{def:Wline}). In the case of gluon operators the Wilson loop is in the adjoint  representation. The rapidity divergences are regularized by the $\delta$-regularization, which consists in suppression of the gluon field in a Wilson line by exponential factor, $A_+(n\sigma+x)\to A_+(n\sigma+x)e^{-\delta|\sigma|}$. The rapidity divergences reveals as $\ln(\delta)$. In this scheme the rapidity renormalization factor is \cite{Echevarria:2012js,Echevarria:2015byo,Vladimirov:2017ksc}
\begin{eqnarray}\label{R=1/S}
R_g\(\vec b,\mu,\zeta|\epsilon,\frac{\delta^+}{p^+}\)=S^{-1/2}\(\vec b|\epsilon,\delta=\frac{\delta^+}{2p^+}\sqrt{\zeta}\).
\end{eqnarray}
The variable $p^+$ is parton momentum \cite{Scimemi:2019gge}, and is required to define the Lorentz invariant scale $\zeta$. Note, that the definition (\ref{R=1/S}) also contains finite at $\delta\to0$ terms, which can be seen as a scheme-dependence. Commonly, the scheme dependence is fixed by condition that no  remnants of the soft factor appear in the hard part of the factorization theorem \cite{Collins:2011zzd,Vladimirov:2017ksc}. Definition (\ref{R=1/S}) satisfies this condition. The UV renormalization factor is taken in $\overline{\text{MS}}$-scheme.
 
The $(\mu,\zeta)$-dependence of gluon TMD distribution is provided by a  pair of evolution equations
\begin{eqnarray}\label{RGE1}
\mu^2 \frac{d}{d\mu^2}\Phi_{g\leftarrow h}(x,\vec b;\mu,\zeta)&=&\frac{\gamma^g(\mu,\zeta)}{2}\Phi_{g\leftarrow h}(x,\vec b;\mu,\zeta),
\\\label{RGE2}
\zeta \frac{d}{d\zeta}\Phi_{g\leftarrow h}(x,\vec b;\mu,\zeta)&=&-\mathcal{D}_g(\mu,\vec b)\Phi_{g\leftarrow h}(x,\vec b;\mu,\zeta).
\end{eqnarray}
These equations are the same for all gluon TMD distributions of  leading twist. The anomalous dimensions are defined via the corresponding renormalization constants and they are known up to three-loop order inclusively \cite{Moch:2005id,Gehrmann:2010ue,Vladimirov:2016dll,Li:2016ctv}. Note, that the renormalization factor $Z_g$ also contains the gluon-field renormalization part, therefore,
\begin{align}
\gamma_G=2 \widehat{AD}(Z_3-Z_g)
\end{align}
where the symbol $\widehat{AD}$ extracts the  coefficient of  $\epsilon^{-1}$ with a pre-factor $n!$ at the $n^{th}$ perturbative order.

Anomalous dimensions $\gamma^g$ and $\mathcal{D}$ satisfy the integrability condition (also known as Collins-Soper equation \cite{Collins:1981va})
\begin{eqnarray}\label{def:inter}
2\mu^2 \frac{d \mathcal{D}_g(\vec b,\mu)}{d\mu^2}=-\zeta \frac{d \gamma^g(\mu,\zeta)}{d\zeta}=\Gamma^g_{\text{cusp}}(\mu),
\end{eqnarray}
where  $\Gamma_{\text{cusp}}$ is anomalous dimension for cusp of two light-like Wilson lines (in the adjoint representation). Due to this equation the expression for $\gamma^g$ can be rewritten in the form
\begin{eqnarray}
\gamma^g(\mu,\zeta)=\Gamma^g_{\text{cusp}}(\mu) \ln\(\frac{\mu^2}{\zeta}\)-\gamma_V^g,
\end{eqnarray}
where $\gamma_V^g$ is anomalous dimension of the vector form factor for gluon. The rapidity anomalous dimension $\mathcal{D}_g$ has not such a simple representation due to the presence of an extra dimensional parameter $\vec b^2$. It generally contains all powers of logarithms $\ln(\mu^2 \vec b^2)$, that at some large values of $\vec b^2$ turns to some non-perturbative function \cite{Scimemi:2016ffw}.

Due to the integrability condition in eq.~(\ref{def:inter}) the system of evolution equations in eq.~(\ref{RGE1},~\ref{RGE2}) has a unique solution:
\begin{eqnarray}\label{def:TMD-evol}
\Phi_{g\leftarrow h}(x,\vec b;\mu_1,\zeta_1)&=&R^g[\vec b;(\mu_1,\zeta_1)\to (\mu_2,\zeta_2)]\Phi_{g\leftarrow h}(x,\vec b;\mu_2,\zeta_2),
\end{eqnarray}
where the TMD renormalization factor reads
\begin{eqnarray}\label{R-def}
R^g[\vec b;(\mu_1,\zeta_1)\to (\mu_2,\zeta_2)]=\exp\Big[\int_P\(\gamma^g(\mu,\zeta)\frac{d\mu}{\mu}-\mathcal{D}_g(\mu,\vec b)\frac{d\zeta}{\zeta}\)\Big].
\end{eqnarray}
Here, $P$ is arbitrary path in $(\mu,\zeta)$-plane connecting $(\mu_1,\zeta_1)$ and $(\mu_2,\zeta_2)$. The eq.~(\ref{R-def}) is in principle independent of the path $P$, however the truncation of the perturbative series makes some choices more preferable, for the detailed discussion see ref.~\cite{Scimemi:2018xaf}. In particular, in sec.~\ref{sec:Results} we use the special practically-convenient path that corresponds to $\zeta$-prescription introduced in \cite{Scimemi:2017etj,Scimemi:2018xaf}. We again stress that the TMD evolution equations and their solution of eq.~(\ref{def:TMD-evol}) do not depend on the polarization, and thus it is exactly same for unpolarized TMDPDF $f_1$ and lpTMDPDF $h_1^\perp$.

\section{Matching coefficient for lpTMDPDF at NNLO}
\label{sec:rrrr}

\subsection{Evaluation of the matching coefficient}

The coefficient function for OPE at twist-2 level can be deduced from the calculation of matrix elements with free parton states with subsequent matching of the result on the desired OPE structures eq.~(\ref{OPE_lpG}). Therefore, the task is naturally split into two steps: the evaluation of parton-matrix element and the matching. This procedure is well-known, see e.g. \cite{Collins:2011zzd,GarciaEchevarria:2011rb,Echevarria:2015usa,Echevarria:2015uaa,Echevarria:2016scs,Gutierrez-Reyes:2018iod}, in this section we present only minimal details and specifics of calculation of lpTMDPDF.

The evaluation of parton matrix elements of the TMD operators at two-loop level is  the most complicated part of the present work. We have used the same technique that was used by our group for NNLO evaluations in refs.~\cite{Echevarria:2015uaa,Echevarria:2016scs,Gutierrez-Reyes:2018iod}, where we refer for extra details. In the case of lpTMDPDF the main complication comes from the rich vector structure, which is reduced to scalar products by projection factor in eq.~(\ref{eq:l-proj}), and the use of unpolarized parton states with momentum $p^\mu=p^+ n^\mu$. In this aspect the current computation is similar to evaluation of the pretzelosity distribution \cite{Gutierrez-Reyes:2018iod} albeit with significantly larger number of loop-integrals. The reduction of integrals to  master integrals and some details of their evaluation is presented in the appendix \ref{app:1}.

The outcome of each diagram at NNLO has a generic form
\begin{align}
\text{diag.}=(\vec b^2)^{2\epsilon}\Big(g_1(x,\epsilon)&+\left(\frac{\delta^+}{p^+}\right)^{\epsilon}g_2(x,\epsilon)+\left(\frac{\delta^+}{p^+}\right)^{-\epsilon}g_3(x,\epsilon)
\\\nn &+\ln\(\frac{\delta^+}{p^+}\) g_4(x,\epsilon)+\ln^2\(\frac{\delta^+}{p^+}\) g_5(x,\epsilon)\Big).
\end{align}
The functions $g_2$ and $g_3$ exactly cancel in the sum of all the diagrams (and this fact can be also traced in the sum of sub-classes of diagrams) because they represent IR divergences. The last two terms represent the rapidity diverging pieces, and thus the functions $g_4$ and $g_5$ are canceled by the rapidity renormalization factor. However, due to the absence of three-order term , the functions $g_5$ cancel in the diagrams. The cancellation of all these pieces provides a check of the calculation.

Summing together the diagrams we obtain the un-subtracted expression for TMDPDF on free-gluon states. Let us introduce the notation for perturbative series
\begin{eqnarray}
h_{1;f\ot f'}^{\perp;\text{unsub.}}(x,\vec b)=\Phi_{f\ot f'}^{\text{unsub.}}(x,\vec b)&=&\sum_{n=1}^\infty a_s^n \Phi_{f\ot f'}^{[n]\text{unsub.}},\qquad 
S(\vec b)=1+\sum_{n=1}^\infty a_s^n S^{[n]},
\end{eqnarray}
where $a_s=g^2/(4\pi)^2$. The tree-order term is zero in the case of lpTMDPDF, 
\begin{eqnarray}
\Phi_{f\ot f'}^{[0]\text{unsub.}}=0,
\end{eqnarray}
which provides many simplifications. In this notation, the expression for the renormalized lpTMDPDF in eq.~(\ref{TMD_ren}) on a parton reads
\begin{align}
\Phi^{[1]}_{f\ot f'}&=\Phi^{[1]\text{unsub.}}_{f \ot f'}
 \label{rentransNNLO}
\\
\Phi^{[2]}_{f\ot f'}&=\Phi^{[2]\text{unsub.}}_{ f\ot f'}-\frac{S^{[1]}\Phi^{[1]\text{unsub.}}_{ f\ot f'}}{2}+Z^{[1]\text{TMD}}_g\Phi^{[1]\text{unsub.}}_{ f\ot f'}.
\end{align}
The expressions for $Z_g^{\text{TMD}}$ is given in ref.~\cite{Echevarria:2016scs}, while the expression for the soft factor in $\delta$-regularization is in ref.~\cite{Echevarria:2015byo}.

Given the values of parton matrix elements we find the coefficient functions matching left- and right-hand sides of
\begin{eqnarray}\label{matching-eqn}
h_{1,g\ot f}^\perp(x,\vec b)=\sum_{f=g,q,\bar q}[\delta^L\!C_{g\ot f'}(\vec b)\otimes f_{1,f'\ot f}](x),
\end{eqnarray}
where $f_{1,f'\ot f}$ is the renormalized parton matrix element for PDF operator eq.~(\ref{def:f1-PDF}), and $\otimes$ is the short-hand notation for Mellin convolution integral eq.~(\ref{OPE_lpG}). Such a relation is valid since the OPE is an operator relation and it is independent of states. 

To solve the matching in eq.~(\ref{matching-eqn}) we need the expression for the collinear matrix elements $f_{1,f'\ot f}$. This calculation is trivial in the actual scheme since there is no Lorenz-invariant scale inside the integrands and all loop-integrals for $f_{1,f'\ot f}$ are zero in dimensional regularization.
 For this reason the loop-corrections to $f_{1,f'\ot f}$ are given by UV renormalization constant only:
\begin{eqnarray}\label{PDF-loops}
f^{[0]}_{1,f\ot f'}(x)=\delta_{ff'}\delta(1-x),\qquad f^{[1]}_{1,f\ot f'}(x)=-\frac{P^{[1]}_{f\ot f'}(x)}{\epsilon},
\end{eqnarray}
where $P^{[1]}$ is the DGLAP evolution kernel at LO.

Denoting the perturbative terms for the matching coefficient as
\begin{eqnarray}
\delta^L\!C_{g\ot f}(x,\vec b)=\sum_{n=1}^\infty a_s^n\delta^L\!C^{[n]}_{g\ot f}(x,\vec b),
\end{eqnarray}
we find from eq. (\ref{matching-eqn},~\ref{PDF-loops}),
\begin{eqnarray}
\delta^L\!C^{[0]}_{g\ot f}(x,\vec b)&=&0,\qquad \delta^L\!C^{[1]}_{g\ot f}(x,\vec b)=h_{g\ot f'}^{\perp [1]}(x,\vec b),
\\\label{qweqeqweqweq}
\delta^L\!C^{[2]}_{g\ot f}(x,\vec b)&=&h_{g\ot f'}^{\perp [2]}(x,\vec b)+\frac{1}{\epsilon}\sum_{f'}[\delta^L\!C^{[1]}_{g\ot f'}(\vec b)\otimes P^{[1]}_{f'\ot f}](x).
\end{eqnarray}
This procedure cancels the collinear poles that are present in the parton matrix elements.
 Note that, the last term in eq.~(\ref{qweqeqweqweq}) requires the evaluation of $\delta^L\!C^{[1]}$ to  order $\sim \epsilon$.

\subsection{Logarithmic part of the  coefficient function}
\label{sec:logs}

The renormalization group equation allows us to write down the coefficients that accompany the scaling logarithms in the coefficient function. We recall that the coefficient function depends on three scales see eq.~(\ref{OPE_lpG}): $\mu$ and $\zeta$ that are inherited from the TMDPDF, and $\tilde\mu$ that is the scale of OPE. The behavior on scales $\mu$ and $\zeta$ is dictated by the TMD evolution equations (\ref{RGE1},~\ref{RGE2}), while the dependence on scale $\tilde \mu$ is canceled by the corresponding dependence of $f_1(x,\tilde \mu)$. The latter is given by the DGLAP equation
\begin{eqnarray}\label{RGE3}
\mu^2 \frac{d}{d\mu^2} f_{1,f\ot h}(x,\mu)&=&\sum_{f'=g,q,\bar q}\int_x^1 \frac{dy}{y}P_{f\ot f'}\(\frac{x}{y},\mu\)f_{1,f'\ot h}(y,\mu).,
\end{eqnarray}
Therefore, at the point $\mu=\tilde \mu$ the coefficient function satisfies the pair of equations
\begin{eqnarray}\label{RGE4}
&&\mu^2\frac{d}{d\mu^2}\delta^L \!C_{g\ot f}(x,\vec b;\mu,\zeta,\mu)
\\&&\nn\qquad= \sum_{f'=g,q,\bar q}\int_x^1 \frac{dy}{y}\delta^L\!C_{g\ot f'}\(\frac{x}{y},\vec b;\mu,\zeta,\mu\)
\(\frac{\gamma_V^g(\mu,\zeta)}{2}\delta_{ff'}\delta(\bar y)-P_{f'\ot f}(y,\mu)\),
\\\label{RGE5}
&&\zeta\frac{d}{d\zeta}\delta^L\!C_{g\ot f}(x,\vec b;\mu,\zeta,\mu)=-\mathcal{D}_g(\mu,\vec b)\delta^L\!C_{g\ot f}(x,\vec b;\mu,\zeta,\mu).
\end{eqnarray}
The solution at NNLO has the simple form
\begin{eqnarray}\label{Log-terms}
\delta^L\!C^{[2]}_{g\ot f}(x,\vec b;\mu,\zeta,\mu)=\(-\frac{1}{2}\mathbf{L}_\mu^2+\mathbf{L}_\mu \mathbf{l}_\zeta\)\delta^L\!C^{(2,1,1)}_{g\ot f}(x)+\mathbf{L}_\mu 
\delta^L\!C^{(2,1,0)}_{g\ot f}(x)+\delta^L\!C^{(2,0,0)}_{g\ot f}(x),
\end{eqnarray}
where
\begin{eqnarray}
\mathbf{L}_\mu=\ln\(\frac{\mu^2\vec b^2}{4e^{-\gamma_E}}\),\qquad \mathbf{l}_\zeta=\ln\(\frac{\mu^2}{\zeta}\).
\end{eqnarray}
The coefficients of logarithms are
\begin{eqnarray}
\delta^L\!C^{(2,1,1)}_{g\ot f}(x)&=&\frac{\Gamma_0^g}{2}\delta^L\!C^{(1,0,0)}_{g\ot f}(x),
\\\nn
\delta^L\!C^{(2,1,0)}_{g\ot f}(x)&=&2\beta_0 \delta^L\!C^{(1,0,0)}_{g\ot f}(x)-\sum_{f'=g,q,\bar q}[\delta^L\!C^{(1,0,0)}_{g\ot f'}\otimes P^{[1]}_{f'\ot f}](x),
\end{eqnarray}
where $\Gamma_0^g=4C_A$ is LO cusp anomalous dimension, $\beta_0=11/3C_A-2/3N_f$ is LO $\beta-$function, and we have used that $\gamma_V^{g[1]}=-2\beta_0$. The explicit expressions for these coefficients are given in the appendix \ref{app:B} for completeness. The finite parts $\delta^L C^{(n;0,0)}$ are presented in the next section. 

In the expressions above we have set $\tilde \mu=\mu$, which is a poor choice. In particular, due to this choice one obtains the double-logarithms in the coefficient function and, as the result, a badly convergent perturbative series. A much better behaved coefficient function can be achieved by distinguishing the scales of evolution and OPE. 
For example, this is realized by applying the  $\zeta$-prescription, which consists in the selection of TMD evolution scales along the null-evolution line in the plane $(\mu,\zeta)$. This line is parameterized as $\zeta=\zeta_\mu(\vec b)$, and it is defined by the boundary condition that it passes through the saddle point of the evolution potential~\cite{Scimemi:2018xaf}. 
The expression for the coefficient function can be obtained by the substitution (here for gluon distributions)
\begin{eqnarray}
\text{in $\zeta$-prescription:}\qquad\mathbf{l}_\zeta=\frac{\mathbf{L}_\mu}{2}-\frac{2\beta_0}{\Gamma_0^g}+\mathcal{O}(a_s).
\end{eqnarray}
The higher order terms and the derivation of this expression can be found in ref.~\cite{Scimemi:2017etj,Scimemi:2018xaf}. The coefficient function in $\zeta$-prescription satisfies DGLAP equation, and thus the remaining scale is the OPE scale $\tilde \mu$. In other words, we have
\begin{eqnarray}
\label{eq:dcl}
\delta^L \!C_{g\ot f}(x,\vec b;\mu,\zeta_\mu(b),\tilde \mu)=\delta^L \!C_{g\ot f}(x,\vec b;\tilde\mu),
\end{eqnarray}
where the logarithmic part has simple form
\begin{eqnarray}
\delta^L \!C^{[2]}_{g\ot f}(x,\vec b;\tilde\mu)=\(\beta_0 \delta^L\!C^{(1,0,0)}_{g\ot f}(x)-\sum_{f'=g,q,\bar q}[\delta^L\!C^{(1,0,0)}_{g\ot f'}\otimes P^{[1]}_{f'\ot f}](x)\)\mathbf{L}_{\tilde \mu}+\delta^L\!C^{(2,0,0)}_{g\ot f}(x).
\end{eqnarray}
The finite part $\delta^L\!C^{(2,0,0)}_{g\ot f}(x)$ remains unaffected. Note that, generally the $\zeta$-prescription also modifies the finite part of NNLO expression, as it is happens e.g. for the unpolarized TMDPDF. 

%%%%%%%%%%%%%%%%%%%%%%%%%%%%%%%%%
\subsection{Finite part of the coefficient function}
\label{sec:finite-parts}
%%%%%%%%%%%%%%%%%%%%%%%%%%%%%%%%%

In this section we present the finite parts of coefficient function $\delta^L\!C$. The NLO expression read
\begin{eqnarray}\label{res:NLO}
\delta^L C^{(1,0,0)}_{g\ot g}(x,\vec b)&=&-C_A
\frac{4(1-x)}{x},
\\\label{res:NLO1}
\delta^L C^{(1,0,0)}_{g\ot q}(x,\vec b)&=&-C_F \frac{4(1-x)}{x},
\end{eqnarray}
where $C_A=N_c(=3)$ and $C_F=(N_c^2-1)/2N_c(=4/3)$ are eigenvalues of quadratic Casimir operators for adjoint and fundamental representations in $SU(N_c)$($SU(3)$) group. The result in eq.~(\ref{res:NLO},~\ref{res:NLO1}) agrees with \cite{Becher:2012yn,Echevarria:2015uaa,Gutierrez-Reyes:2017glx}. The full $\epsilon$-dependent NLO expressions are presented in \cite{Gutierrez-Reyes:2017glx}. The NNLO expressions are \footnote{{We thank Hua Xing Zhu et al. for pointing out an error of the final result  in the previous version of the paper.}}
\begin{align}
\nn
\delta^L \!C_{g\ot g}^{(2;0,0)}(x)&={C_A^2\Big [
-16 \frac{1-x}{x}\(\Li_2(x)-\ln x\)+\frac{124}{3}\ln x+\(\frac{148}{9}+20\zeta_2\)\frac{1-x}{x}}\nn\\
&{-8\ln^2 x-\frac{100}{9}(1-x)-\frac{4}{9}x(11x-14)
\Big ]}\nn
\\%\nn
\label{C2gg} 
&+C_F N_f\cdot 4\Big[\ln^2x-2\frac{(1-x)^3}{x} \Big]
+C_A N_f\cdot \frac{4}{9}\Big[17\frac{1-x}{x}+1-3x-x^2+6\ln x\Big],
\\
\nn
\delta^L \!C^{(2;0,0)}_{g\ot  q}(x)&=C_F(C_F-C_A)\Big[8\frac{1-x}{x}(\ln(1-x)+\ln^2(1-x))-20\ln x+4\ln^2x+8(1-x)\Big]
\\\nn &+C_FC_A\Big[{16\frac{1-x}{x}\(\frac{11}{18}+\frac{5}{4}\zeta_2-\frac{\ln(1-x)}{3}-\Li_2(x)\)}
\\%\nn
 &{+4\frac{\ln x}{x} \(4+5x-x\ln x\)}\Big]+C_FN_f\cdot \frac{16}{9}\frac{1-x}{x}[2+3\ln(1-x)],
%\label{eq:NNLO1}
\label{C2gq}
\end{align}
where $N_f$ is the number of active quark flavors. These expressions is the main result of this work. {These results have been recently obtained in ref.~\cite{Luo:2019bmw} with an independent calculation using the exponential regulator of ref.~\cite{Li:2016axz} to regularize rapidity divergences. We find full agreement with the final results presented.}

\section{lpTMDPDF at NNLO and its contribution Higgs production}
\label{sec:Results}

The lpTMDPDF and the  unpolarized gluon TMDPDF use to be present at the same time in many processes. A particularly important place to study the effect of lpTMDPDF is the Higgs production in hadron-hadron collision. In this case the dominating channel for Higgs production is gluon-gluon fusion via the top-quark loop \cite{Ellis:1975ap}, which can be written via an effective interaction term in the Lagrangian \cite{Shifman:1979eb}
\begin{eqnarray}\label{ggH}
\mathcal{L}_{ggH}=\frac{a_s(\mu)C_t(\mu)}{3v}\,F^A_{\mu\nu}F^{A,\mu\nu}H,
\end{eqnarray}
where $H$ is the Higgs field, $F_{\mu\nu}$ is the gluon field strength tensor, and $v$ is the Higgs vacuum expectation value. The effective coupling constant at NNLO is derived in \cite{Kramer:1996iq,Chetyrkin:1997iv}. Using  the effective vertex in eq.~(\ref{ggH}) one can derive the TMD factorization theorem for Higgs production following the same steps as in the Drell-Yan case (see  refs.~\cite{Georgi:1977gs,Ahrens:2008nc,Ravindran:2003um,Anastasiou:2005qj}). The resulting expression is 
\begin{eqnarray}\label{Higgs-xSec}
\frac{d\sigma}{dy d^2\vec q_T}&=&\frac{2\sigma_0(\mu)}{\pi}C_t^2(\mu)U(\mu,-\mu)|C_H(-m_H^2,-\mu^2)|^2
\\\nn && \int\frac{d^2\vec b}{4\pi}e^{i(\vec b\vec q_T)}\Phi^{\mu\nu}_{g\ot h_1}(x_1,\vec b;\mu,\zeta_1)\Phi^{\mu\nu}_{g\ot h_2}(x_2,\vec b;\mu,\zeta_2),
\end{eqnarray}
where $y$ is the Higgs rapidity and $x_{1,2}=\sqrt{(m_H^2+\vec q_T^2)/s}e^{\pm y}$. The function $C_H$ is the gluon scalar form-factor (the NNLO expression can be found in \cite{Gehrmann:2010ue}), $U$ is the ``$\pi^2$-resummation'' exponent \cite{Ahrens:2008qu} and the TMD distributions $\Phi^{\mu\nu}$ are defined in eq.~(\ref{def:TMD_OP_G}). For a more accurate and detailed definition we refer to ref.~\cite{Ahrens:2008nc}. The scale $\mu$ is of the order of the hard scale, $m_H$ in this case, and $\zeta_1\zeta_2=m_H^4$. 

With the decomposition in eq.~(\ref{TMD_G_dec}) the product of TMD distributions turns into
\begin{align}\label{PHI-PHI}
\Phi^{\mu\nu}_{g\ot h_1}(x_1,\vec b)\Phi^{\mu\nu}_{g\ot h_2}(x_2,\vec b)=\frac{1}{2}\(f_{1,g\ot h_1}(x_1,\vec b)f_{1,g\ot h_2}(x_2,\vec b)
+h^\perp_{1,g\ot h_1}(x_1,\vec b)h^\perp_{1,g\ot h_2}(x_2,\vec b)\) .\,
\end{align}
Therefore, for a consistent phenomenological application of this formula  one should consider $f_1$ and $h_1^\perp$ at the same perturbative order. 
The perturbative inputs up to NNLO are  reported in tab.~\ref{tab:pert}.

%%%%%%%%%%%%%%%%%%%
\begin{table}[h]
\begin{center}
\begin{tabular}[h]{|c||c|c||c|c|c||c|c|}\hline
Function & $H$ &  $C_{g\ot f}$,  $\delta^L C_{g\ot f}$ & $\Gamma_{\text{cusp}}$ & $\mathcal{D}$ & $\gamma_F$ & $\alpha_s$ running & PDF evolution
\\\hline 
NLO  & $\alpha_s$ & $\alpha_s$ & $\alpha_s^2$ & \specialcellcenter{$\alpha_s$\\ resummed} & $\alpha_s^2$ & \multicolumn{2}{|c||}{\specialcellcenter{NLO provided by \\ NNPDF3.1~\cite{Ball:2017nwa}} }
\\\hline
NNLO  & $\alpha_s^2$ & $\alpha_s^2$ & $\alpha_s^3$ & \specialcellcenter{$\alpha_s^2$\\ resummed} & $\alpha_s^3$ & \multicolumn{2}{|c||}{\specialcellcenter{NNLO provided by \\ NNPDF3.1~\cite{Ball:2017nwa}} }
\\\hline
\end{tabular}
\end{center}
\caption{\label{tab:pert} Summary of perturbative orders used  for each part of the cross section. The symbol $H$ stands for the first line of eq.~(\ref{Higgs-xSec}).
}
\end{table}
%%%%%%%%%%%%%%%%%%%%%

 It is interesting to mention that if the Higgs boson were a pseudo-scalar particle, then the main change in the structure of cross-section in eq.~(\ref{Higgs-xSec}) would be a sign of $h^\perp_{1}h^\perp_{1}$ term in eq.~(\ref{PHI-PHI}).  
In this case, the expressions for perturbative corrections in $C_t$ and $C_H$ are also changed although their LO remains the same \cite{Boer:2011kf}.

In order to study the numerical impact of our result,  the NLO and NNLO matching for lpTMDPDF together with the cross-section in eq.~(\ref{Higgs-xSec}) have been added to \texttt{artemide}~\cite{web}. 
The non-perturbative parts of gluon TMD distributions  and gluon rapidity anomalous dimension are unknown, and nowadays the data are not sufficient to fix it.
In order to provide some value for a cross section we use the inputs in eq.~(\ref{h_ansatz}-\ref{eq:lig}) with $f_{NP}=f_{1NP}=h^\perp_{1NP}$, where $f_{NP}$ is the non-perturbative  function for quarks extracted from a fit of Drell-Yan and Z-boson production data using   \texttt{artemide2.01}. The details of this fit have been illustrated in ref.~\cite{Scimemi:2017etj,Bertone:2019nxa}, and this version of the code takes into account the improvements coming from ref.~\cite{Vladimirov:2019bfa}. The TMD evolution kernel for gluons should be also provided by a non-perturbative part at large value of $b$, whose precise analytical form is given in  \cite{Bertone:2019nxa}. The perturbative calculable parts of the evolution kernel differ in quark and gluon case (at the order that we work)  by the Casimir scaling factor
$C_A/C_F$. Here we have assumed the same scaling for the un-calculable non-perturbative pieces of the evolution kernel. The error  band of our prediction come  from  scale variations of a factor of 2, consistently with  $\zeta$-prescription  \cite{Scimemi:2018xaf}.

\begin{figure}[t]
\centerline{
\includegraphics[width =0.5\textwidth]{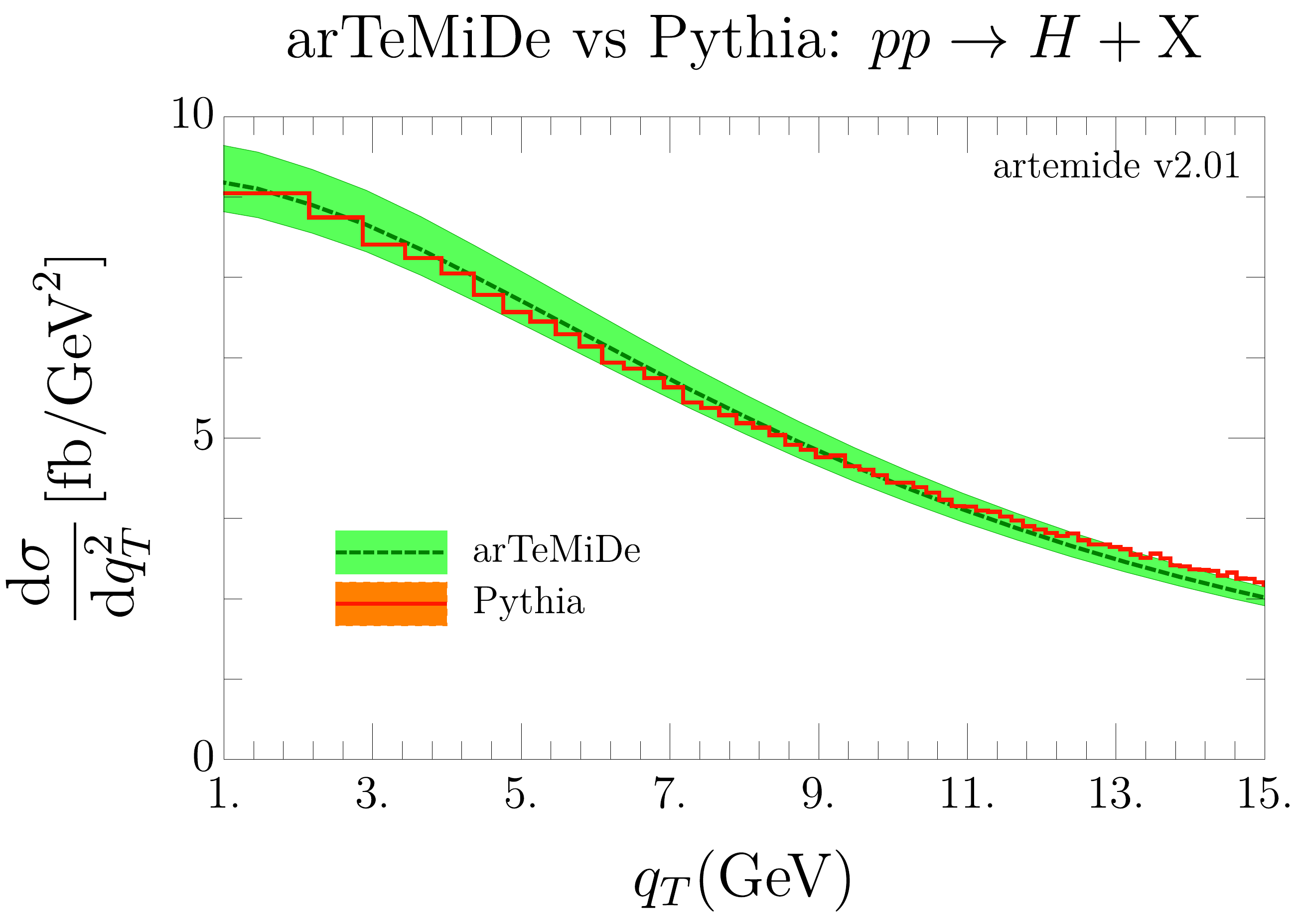}}
\caption{The cross section in eq.~(\ref{Higgs-xSec}) integrated over all rapidity range with \texttt{artemide2.01} at NNLO and PYTHIA. The errors of PYTHIA are included, although not clearly visible.
\label{pythia} The shaded area shows the variation band in $\tilde \mu$, see eq.~(\ref{eq:dcl}).}
\end{figure}

 In order to check the  viability of the model assumptions we have compared the cross section in eq.~(\ref{Higgs-xSec}), integrated in rapidity,  with 
PYTHIA \cite{Sjostrand:2006za,Sjostrand:2007gs}. The agreement  of our prediction  at NNLO and PYTHIA is shown in fig.~\ref{pythia} and it is extremely good in the range of $q_T$ where the  TMD factorization theorem is expected to hold. In that figure we have also included the error provided by  PYTHIA, although it is not clearly visible.

%%%%%%%%%%%%%%%%%%%%%%%%%%%%%%%%%%%%%%%%
%%%%%%%%%%%%%%%%%%%%%%%%%%%%%%%%%%%%%%%

\begin{figure}[t]
\centerline{
\includegraphics[width =0.48\textwidth]{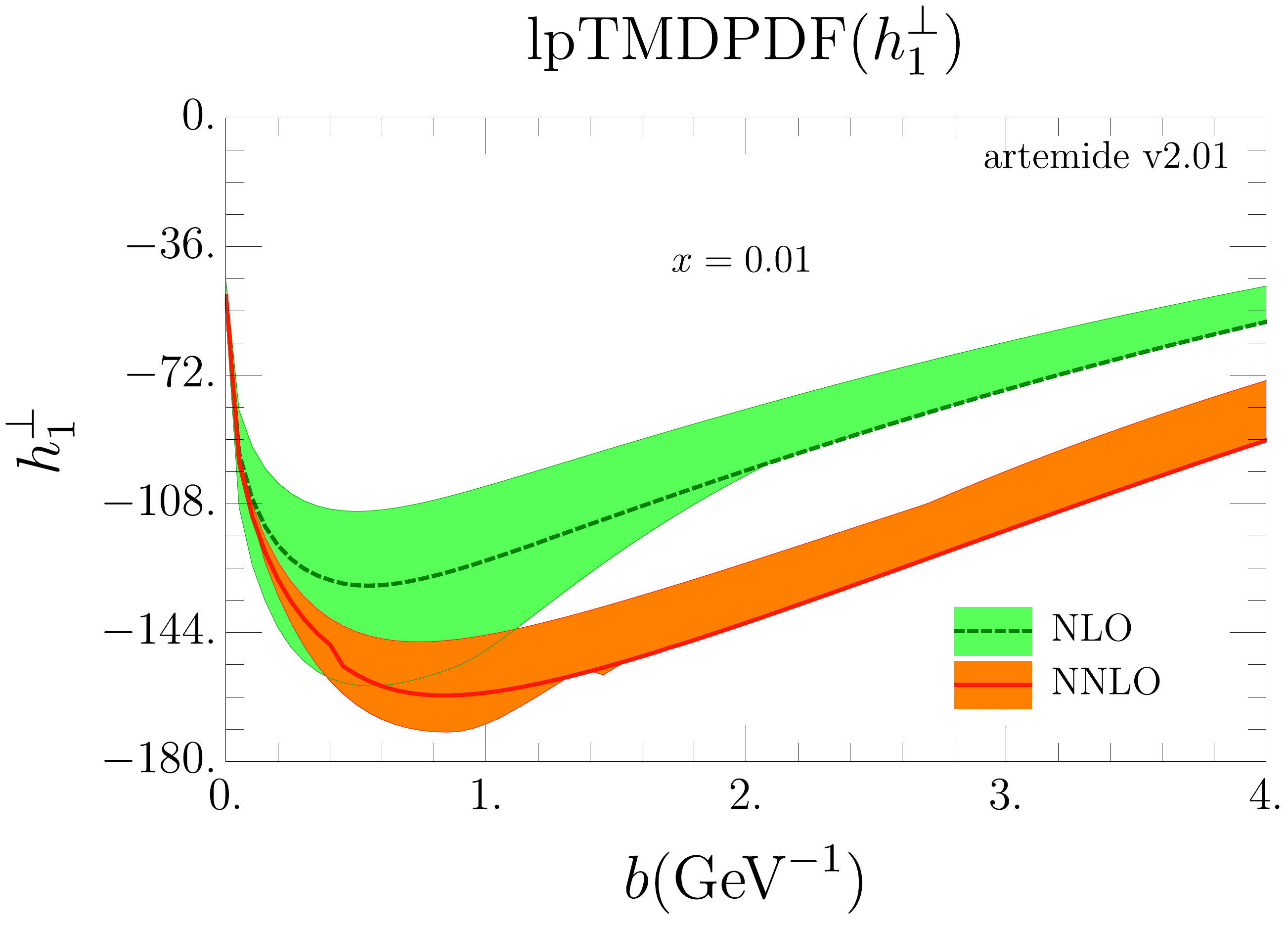}
~~
\includegraphics[width =0.48\textwidth]{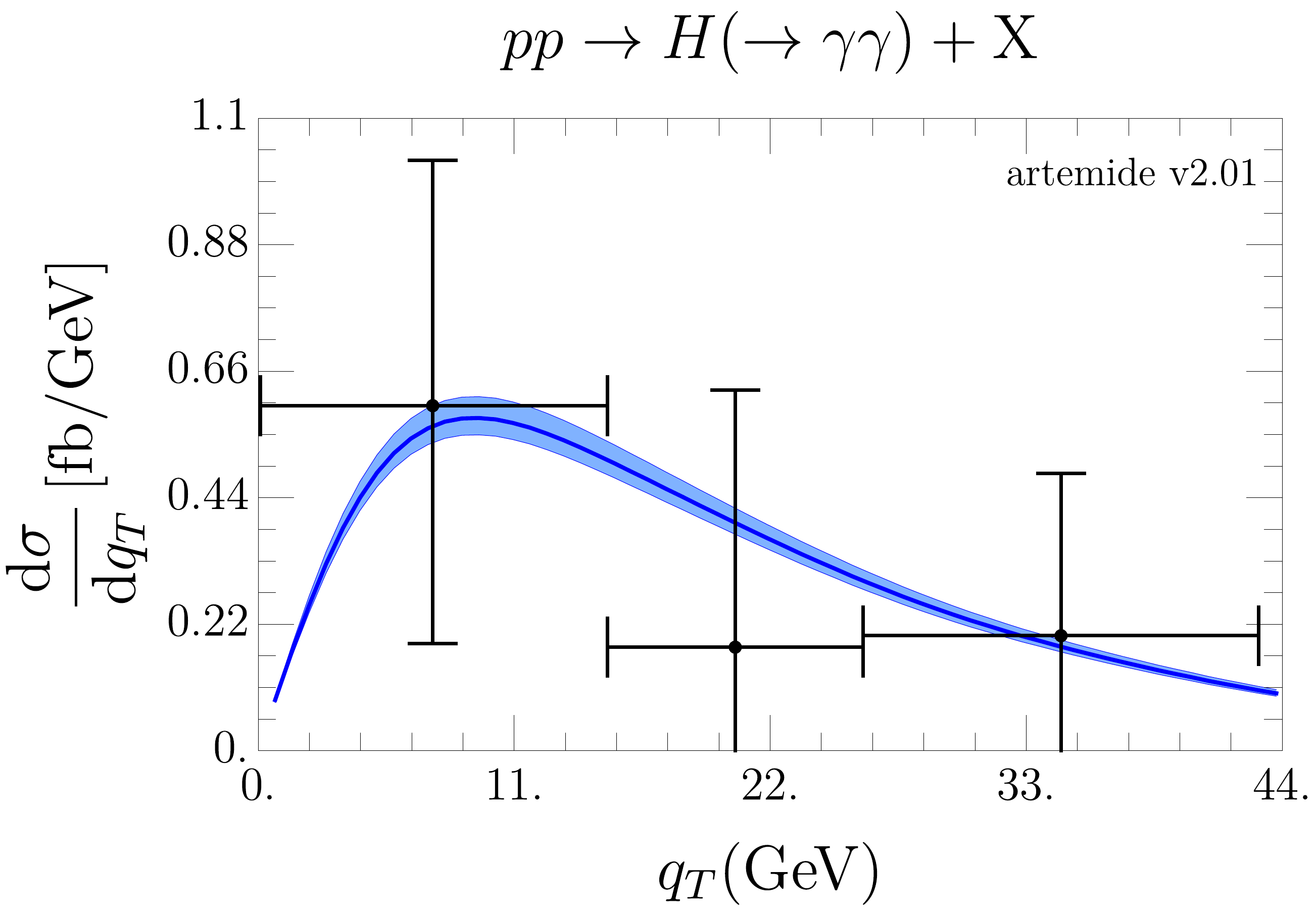}
}
\caption{
(left) The lpTMDPDF, eq.~(\ref{h_ansatz}-\ref{eq:lig}), as a function of $\vec b$ at $x=0.01$. The shaded area shows the variation band in $\tilde \mu$, see eq.~(\ref{eq:dcl}).
(right) Comparison of Higgs-production cross-section with variation band to the measurement presented in \cite{Khachatryan:2015rxa} by CMS collaboration. 
}
\label{fig:lpA}
\end{figure}

% (see fig.~\ref{fig:lpA}(left), for comparison of the theory and experimental precisions). Therefore, we have used the following assumptions. The non-perturbative functions $f_{\text{NP}}$ in eq.~(\ref{h_ansatz}) have been taken the same for both $f_1$ and $h_1^\perp$ with the same values as it was extracted in the quark case in \cite{Scimemi:2017etj} (with unpolarized collinear gluon PDF taken from NNPDF3.1 set \cite{Ball:2017nwa}). It should give a reasonable estimation for the sizes of non-perturbative effects, while the main numerical difference comes from the values of matching coefficients. The rapidity anomalous dimension is taken the same as in extracted \cite{Bertone:2019nxa} with additional rescaling by factor $C_A/C_F$ (Casimir scaling).
In fig.~\ref{fig:lpA} (left) we have plotted  lpTMDPDF, eq.~(\ref{h_ansatz}-\ref{eq:lig}), as a function of $\vec b$ at $x=0.01$ at NLO and at NNLO. 
 The NNLO includes the perturbative correction to the first non-trivial order (which is NLO). This correction appears to be large, say almost a factor 2.
%Since the tree level matching of the lpTMDPDF is null, and the actual LO is $\sim a_s^1$, lpTMDPDF is expected to receive a relatively large correction at order $a_s^2$. \blue{Some examples of cases similar to this one are given in e.g.~\cite{Bozzi:2007pn,Scimemi:2017etj}}. This fact is confirmed in fig.~\ref{fig:lpA}(right). The difference between NLO and NNLO is practically of order 2 at $x=0.01$. 
The bands show the sensitivity of the  distribution to the change of the  OPE scale $\tilde \mu\to c_4\tilde \mu$ with $c_4\in (0.5,2)$, see eq.~(\ref{eq:dcl}). The relative size of the band decreases between NLO and NNLO.
Altogether, this figure points to the fact that the lpTMDPDF effects could have been underestimated up to now. 

The experimental data on the Higgs differential cross section are still affected by big errors. For a demonstration we have considered the cross section in eq.~(\ref{Higgs-xSec}) measured at CMS collaboration, where the rapidity is integrated in the interval indicated by that experiment~\cite{Khachatryan:2015rxa}.  Because the experimental cross section just uses the data from one particular decay of the Higgs boson we have normalized our cross section with the experimental one integrating in the interval of transverse momenta shown  in fig.~\ref{fig:lpA} (right). From this figure it is clear that  currently the data are not sensitive to the TMD structures.

\begin{figure}[t]
\centerline{
\includegraphics[width =0.48\textwidth]{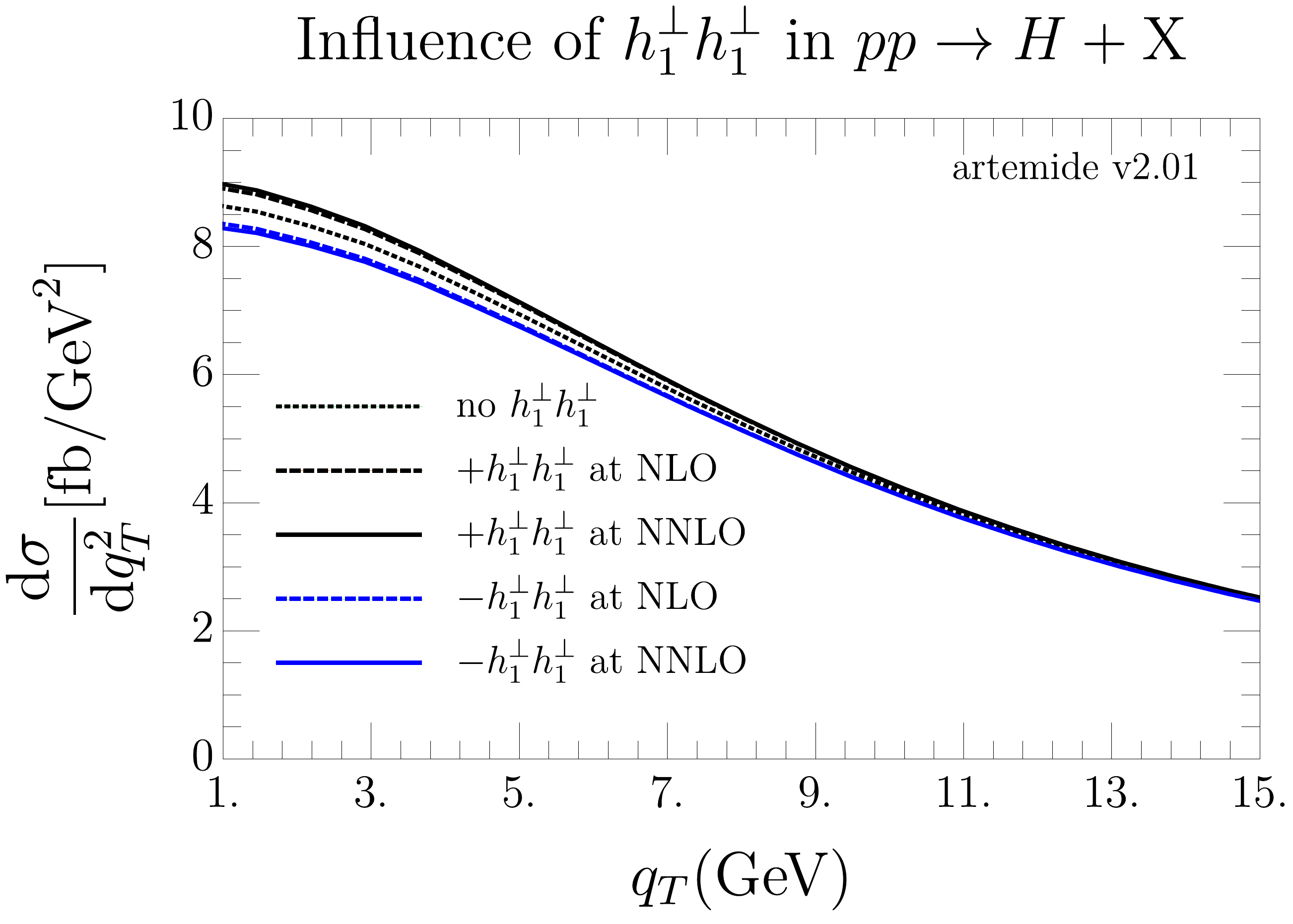}
~~
\includegraphics[width =0.48\textwidth]{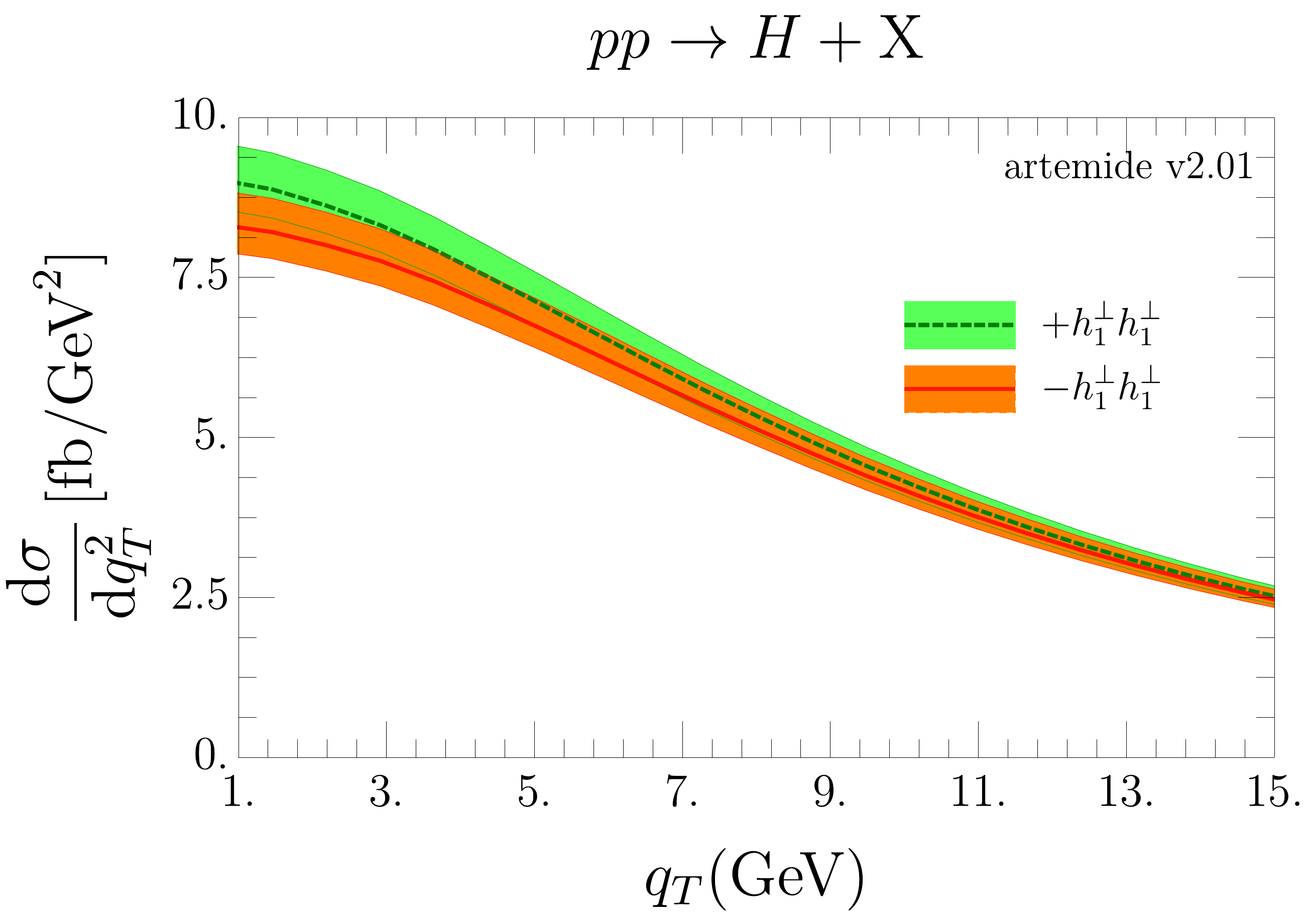}}
\caption{Cross section for Higgs production including linearly polarized gluon effects at different orders. 
(left) The motion of center lines of cross-section integrated over all rapidities at different perturbative orders for lpTMDPDF. 
The black (blue) lines correspond the case of positive (negative) contribution for $h_1^\perp h_1^\perp$-term in eq.~(\ref{PHI-PHI}). 
 (right) The scale-variation band for the cross section at NNLO. The parity even (odd) Higgs case is represented with a green (orange) band. In both figures the center of mass energy is set as in~\cite{Khachatryan:2015rxa} and rapidity is integrated over its complete range.}
\label{fig:xSec1}
\end{figure}

In the Higgs production cross-section the  lpTMDPDF mainly affects the low-$q_T$ region, as it is demonstrated in fig.~\ref{fig:xSec1}. 
Practically, the lpTMDPDF can be distinguished from  the unpolarized TMDPDF at $q_T\lesssim 5$-8GeV, where it modifies the values of cross-section by about $5\%$. Such value of variation band is typical for NNLO approximation, see e.g.~\cite{Cruz-Martinez:2018rod}. In fig.~\ref{fig:xSec1} (right) we compare the  NNLO cross sections the size of the variation band, which is the maximum deviation value obtained from the variation of all three scales (in $\zeta$-prescription) by factors $c_i\in(0.5,2)$ \cite{Scimemi:2018xaf}. 
The variation band is of the order of few percents and the main contribution to it is the $\mu$-band (the scale between hard part and the TMD-evolution factor). Nowadays, these factors can be pushed to N$^3$LO reducing the variation band further, if necessary.

\begin{figure}[t]
  \centerline{\includegraphics[width =0.47\textwidth]{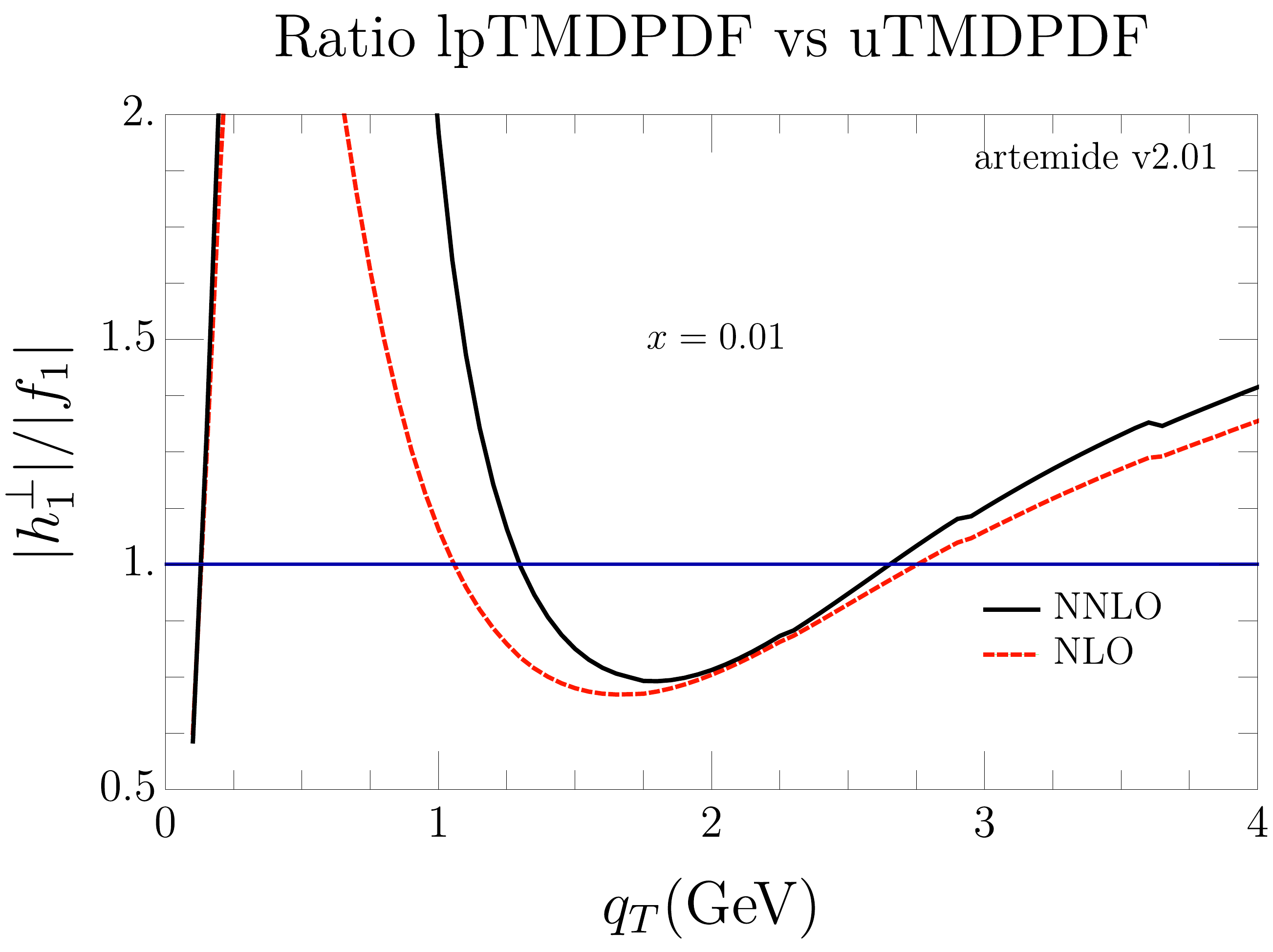}
  \hspace{1cm}\includegraphics[width =0.47\textwidth]{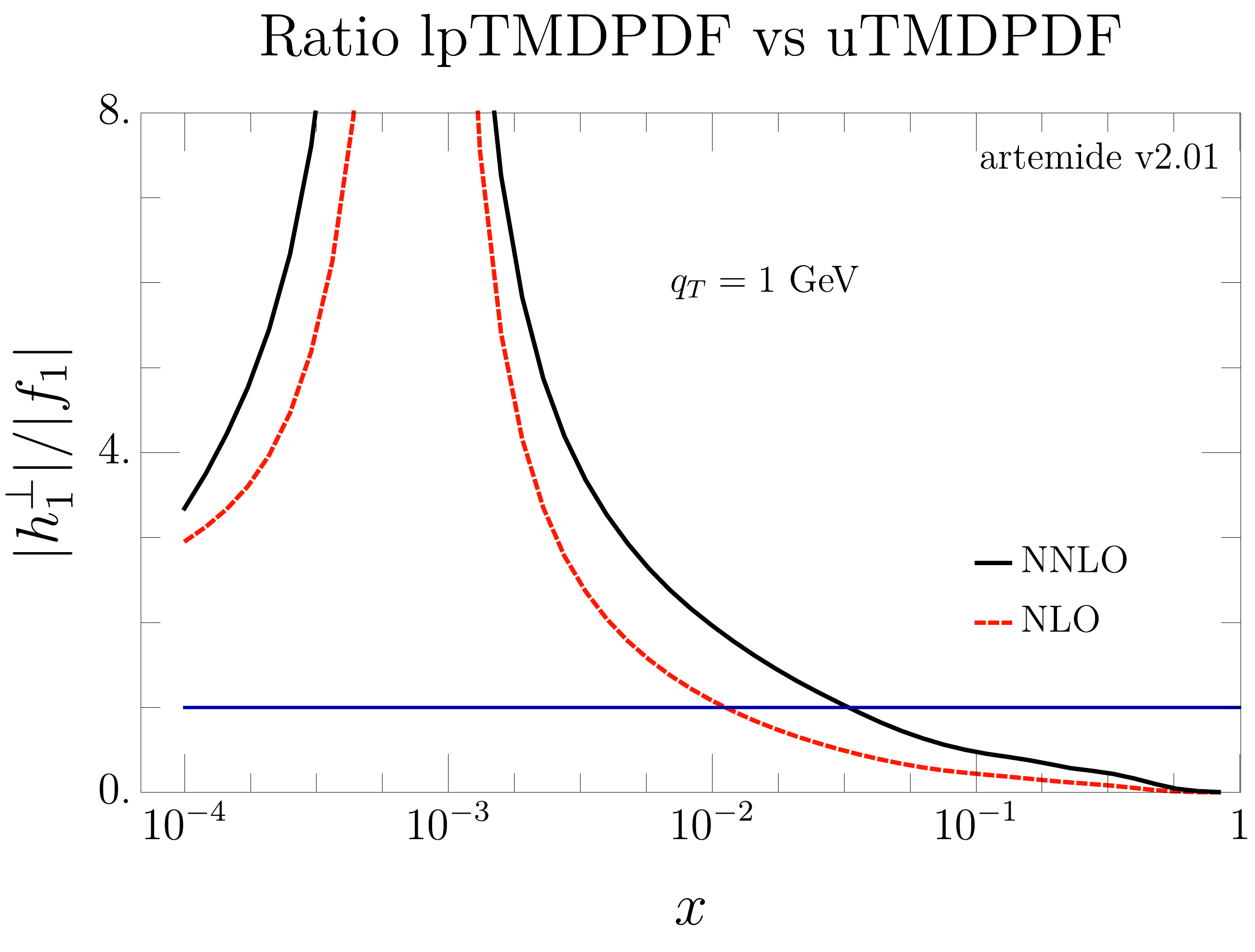}}
  \caption{Ratio of linearly polarized and unpolarized gTMD to  check   eq.~(\ref{eq:positivitybound}) as a function of $q_T$ at  fixed $x=0.01$ (left) and as a function of  $x$ at fixed $q_T=1$ GeV (right).
 }
  \label{fig:ratio2}
\end{figure}

Finally, we comment on the positivity relation formulated in ref.~\cite{Mulders:2000sh}:
\begin{eqnarray}\label{inequality}
|f_1(x,\vec q_T)|-|h_1^\perp(x,\vec q_T)|>0.
\end{eqnarray}
This relation is a consequence of positive definiteness of the gluon-polarization matrix in a free theory, and certainty hold at LO. However, it does not need to be accomplished at higher order in perturbation theory. 
The  positivity bound is formulated in  momentum space, whereas all perturbative calculation are performed in coordinate space. This causes an additional problem since the Hankel transform of a positive function is not necessary a positive function. Within our model we have checked that it is easy to get a violation of this bound, for any fixed value of $x$ and $q_T$. Typically, the violation happens in the vicinity of sign change point of $f_1$ (note, that our realization of $f_1$ is positive-definite in $\vec b$-space). Outside of this point the inequality in eq.~(\ref{inequality}) is respected. The situation is exemplified in fig.~\ref{fig:ratio2}, where we plot the ratio of $|f_1|/|h_1^\perp|$ at different values of $q_T$ with fixed $x$ (left) and viceversa (right). We also note that the positions of zeros in TMDPDFs strongly depends on the non-perturbative input. In particular, selecting some appropriate model one can, possibly, remove the zero from unpolarized TMDPDF, or fix positions of zeros  equal in both gluon TMDPDFs. In other words eq.~(\ref{inequality}) can be used as a serious constraint on non-perturbative part of the TMD distributions. However, we do not see enough theoretical justification for such an approach at the moment.

We have also observed that the ratio $|h_1^\perp|/|f_1|$ tends to saturate at smaller values of $x$ as  it is suggested  f.i. by~\cite{McLerran:1993ka}. Then for extreme small values of $x\sim 10^{-4}$ it is violated again. However, such values can be outside the  applicability region of our calculation since the perturbative expressions for $f_1$ \cite{Echevarria:2016scs} and $h_1^\perp$ (\ref{C2gg},~\ref{C2gq}) have contributions $\sim a_s^{n+1}\ln^n(x)/x$ that should be resummed for a proper comparison.

\section{Conclusions}

The gluon transverse momentum dependent parton distribution function (lpTMDPDF) typically accompanies unpolarized gluon TMDPDF within a TMD factorized cross-section. A good example is the factorization formula for the Higgs-production cross-section, where these distributions enter in a plain sum. For this reason, both distributions should be considered at the same order of perturbative accuracy. We have calculated the $a_s^2$-part (NNLO) for the matching coefficient of lpTMDPDF to twist-2 collinear distributions, which is the main result of this paper. 
Thanks to this calculation, lpTMDPDF can be considered at the same level of theoretical accuracy as the unpolarized gluon TMDPDF \cite{Gehrmann:2014yya,Echevarria:2016scs}. 
The corresponding formulas are collected in sec.\ref{sec:logs},~\ref{sec:finite-parts}. They are also attached to the publication in the form of \textit{Mathematica}-notebook. The module for the numerical evaluation of lpTMDPDF is added to the \texttt{artemide} package that can be downloaded from \cite{web}.

The impact of NNLO correction for lpTMDPDF is very significant and practically doubles the value of the function for moderate $\vec b$. This fact should not be considered much surprising given that LO term ($a_s^0$-term) for lpTMDPDF vanishes and the correction that we provide is the one to the first non-null order.
 The relevance of this effect in the Higgs cross section has been discussed in sec.~\ref{sec:Results}  and it is resumed in figs.~\ref{fig:lpA}-\ref{fig:xSec1}. Unfortunately, at the moment we have not a reliable model for the non-perturbative part of the gluon TMD distribution, and in this work, we have adapted values for distributions extracted in refs. \cite{Scimemi:2017etj,Bertone:2019nxa}. A more detailed study on the non-perturbative part of the gluon TMDPDF is certainly worth in the future. Surprisingly, the model built by us agrees with PYTHIA prediction  for low $q_T$ values, which are the relevant ones for  TMD studies.

In several papers, it has been suggested that unpolarized and linearly polarized gluon TMDPDFs can be measured in association with heavy-quark production~\cite{Mukherjee:2016cjw,Boer:2017xpy,Efremov:2017iwh,Efremov:2018myn,Kishore:2018ugo,Echevarria:2019ynx,Marquet:2017xwy}. We leave an analysis of these processes for future work because at the moment we miss a full factorization theorem for each of these cases. Nevertheless, the consistency of data with the factorization hypothesis can always be checked with the result provided in this work.

\section*{Acknowledgements}
D.G.R., S.L.G. and  I.S. are supported by the Spanish MECD grant FPA2016-75654-C2-2-P. 
This project has received funding from the European Union Horizon 2020 research and innovation program under grant agreement No 824093 (STRONG-2020). D.G.R. acknowledges the support of the Universidad Complutense de Madrid through the predoctoral grant CT17/17-CT18/17. S.L.G. is supported by the Austrian Science Fund FWF under the Doctoral Program W1252-N27 Particles and Interactions.
\appendix

\section{Relevant set of master integrals for linearly polarized gluon TMD}
\label{app:1}

Three different types of diagrams arise in the calculation of the unsubstracted TMDPDF matrix element for linearly polarized gluons and the can be addressed on the basis that the exchanged gluons are pure-virtual, virtual-real or real-real. The pure-virtual diagrams, are zero in the dimensional regularization due to the absence of a Lorentz-invariant scale in our scheme of calculation. 
The  virtual-real and real-real diagrams  have respectively one and two cut propagators and should be computed directly. 
The calculation of these two types of diagrams is analogous to the calculation made in ref.~\cite{Echevarria:2015usa,Echevarria:2016scs} for the case of unpolarized TMDPDF, albeit with a different Lorentz structure. The main difference and difficulty comes from the term proportional to $b^\mu b^\nu$. The contraction of this term with the projectors generates terms in the numerator as $(\boldsymbol{bq})^2$ (where $\vec q$ is a loop-momentum), making the evaluation of the diagrams involved.

For virtual-real diagrams this difficulty can be by-passed by calculating separately virtual subdiagrams. This approach allows to contract the projector only with the real loop-momentum, simplifying the calculation of integrals. For real-real integrals no subdiagrams can be calculated. A set of master integrals in which these diagrams can be decomposed was developed in \cite{Echevarria:2016scs}. In this appendix we present the decomposition of the  master integrals original for this work.

A general master integral can be  written as
\begin{align}
F_{abcd}[R]=(2\pi)^2\int \frac{d^{d-1}k \, d^{d-1}l}{(2\pi)^{2d}} \frac{R\, e^{i(\boldsymbol {kb})}e^{i(\boldsymbol {lb})}\delta(k^2)\theta(-k^-)\delta(l^2)\theta(-l^-)}{[(l+p)^2]^a [(k+p)^2]^b[(k+l+p)^2]^c[(k+l)^2]^d}\ ,
\end{align}
where $R=\{1,(\boldsymbol{kb})^2,(\boldsymbol{kb})(\boldsymbol{lb}),(\boldsymbol{lb})^2\}$. The bold font denotes the scalar product of transverse components only with Euclidian metric. The components $k^+$ and $l^+$ can be integrated with the help of the introduction of a delta function
\begin{align}
1=\int_{-\infty}^\infty d\eta\, p^+ \delta\((1-\eta)p^++l^+\)
\end{align}
and they do not enter in the loop-integration (indicated by a $d-1$ integral). 

The integrals with $R=1$, $F_{abcd}[1]\equiv F_{abcd}$ are presented in the Appendix C of \cite{Echevarria:2016scs}. In that case, the sum of the indices  $abcd$ of the integral is 2. In the present calculation, the new integrals with $R\neq1$ and the  sum of the indices $abcd$ is 3. Some of the new integrals can be expressed as a combination of older results, 
\begin{align}
F_{0210}[(\vec {kb})^2]/\vec B=&2\((1+2\epsilon)(x-\eta)-\frac{\epsilon(1-2\epsilon)}{1+\epsilon}\frac{1-\eta}{x}\)F_{0110}-\frac{2(1-2\epsilon)}{1+\epsilon}(1-x)F_{0020},
\label{F0210kb2}
\\
F_{0210}[(\vec {kb})(\vec{lb})]/\vec B=&\frac{2(1-2\epsilon)}{1+\epsilon}\frac{1-\eta}{1+x-\eta}\(\epsilon \frac{1-\eta}{x}F_{0110}-(1+\epsilon)(\eta-x)F_{0110}+(1-x)F_{0020}\)\nn\\
&+2(1+x-\eta)F_{(-1)210}+2\eta F_{0110},
\label{F0210kblb}
\\
F_{0210}[(\vec{lb})^2]/\vec B=&\frac{x}{(1+x-\eta)^2}\Big(\frac{2(1-2\epsilon)}{1+\epsilon}(x(\eta-x)-(1+\epsilon)(1-\eta))F_{0110}\nn\\
&+\epsilon \frac{2(1-2\epsilon)}{1+\epsilon}x(1-x)F_{0020}-2(1-2\epsilon)(1-\eta)^2F_{0110}\Big)-4(1-\eta)F_{(-1)210}\nn\\
&-\frac{1}{(1+x-\eta)^2}\(\frac{2\epsilon(1-2\epsilon)}{1+\epsilon}\frac{(1-\eta)^3}{x}-2\eta(1-2\epsilon)(1-\eta)^2\)F_{0110}\nn\\
&-\frac{2(1-2\epsilon)}{1+\epsilon}\frac{(1-\eta)^2}{(1+x-\eta)^2}(1-x)F_{0020},
\label{F0210lb2}
\\
F_{0120}[(\vec{kb})^2]/\vec B=&\(4(x-\eta)+\frac{2(1-2\epsilon)}{1+\epsilon}\frac{1-x}{x}(1+x-\eta)\)F_{0020}\nn\\
&+\epsilon\frac{2(1-2\epsilon)}{1+\epsilon}\frac{1-\eta}{x^2}(1+x-\eta)F_{0110},
\label{F0120kb2}
\\
F_{0120}[(\vec{kb})(\vec{lb})]/\vec B=&-\frac{2(1-2\epsilon)}{1+\epsilon}\frac{1-\eta}{x}\(\frac{\epsilon (1-\eta)}{x}F_{0110}+(1-x)F_{0020}\)\nn\\
&-2F_{0110}+2(1+x-\eta)F_{(-1)210}+2\eta F_{0020},
\label{F0120kblb}
\end{align}

\begin{align}
F_{0120}[(\vec{lb})^2]/\vec B=&\frac{2(1-2\epsilon)}{1+\epsilon}\frac{(1-\eta)^2}{x(1+x-\eta)}\(\epsilon \frac{1-\eta}{x}F_{0110}+(1+\epsilon)\frac{x}{1-\eta}F_{0110}+(1-x)F_{0020}\)\nn\\
&-4(1-\eta)F_{(-1)210},
\label{F0120lb2}
\\
F_{1020}[(\vec{kb})^2]/\vec B=&\frac{2(1-2\epsilon)}{1+\epsilon}\frac{(\eta-x)^2}{x\eta}\(\epsilon \frac{\eta-x}{x}F_{1010}+(1+\epsilon)\frac{x}{\eta-x}F_{1010}+(1-x)F_{0020}\)\nn\\
&+4(x-\eta)F_{1(-1)20},
\label{F1020kb2}
\\
F_{1020}[(\vec{kb})(\vec{lb})]/\vec B=&-\frac{2(1-2\epsilon)}{1+\epsilon}\frac{\eta-x}{x}\(\epsilon \frac{\eta-x}{x}F_{1010}+(1-x)F_{0020}\)\nn\\
&-2F_{1010}+2(1+x-\eta)F_{0020}+2\eta F_{1(-1)20},
\label{F1020kblb}
\end{align}
\begin{align}
F_{1020}[(\vec{lb})^2]/\vec B=&\frac{2(1-2\epsilon)}{1+\epsilon}\frac{\eta}{x}\(\epsilon \frac{\eta-x}{x}F_{1010}+(1-x)F_{0020}\)\nn\\
&-4(1-\eta)F_{0020},
\label{F1020lb2}
\\
F_{0021}[(\vec{kb})^2]/\vec B=&-\frac{2(1-2\epsilon)}{1-\epsilon}\frac{\eta-x}{1-x}\((1+x-\eta)F_{0020}-(\eta-x)F_{0011}\)\nn\\
&+4(x-\eta)\(F_{0011}-F_{0020}-F_{(-1)021}\),
\label{F0021kb2}
\\
F_{0021}[(\vec{kb})(\vec{lb})]/\vec B=&\frac{2(1-2\epsilon)}{1-\epsilon}\frac{(\eta-x)(1-\eta)}{1-x}\(F_{0020}+F_{0011}\)+2(1+x-2\eta)F_{(-1)021}\nn\\
&-2(1-\eta)\(F_{0011}-F_{0020}\)-2F_{0020},
\label{F0021kblb}
\\
F_{0021}[(\vec{lb})^2]/\vec B=&-\frac{2(1-2\epsilon)}{1-\epsilon}\frac{1-\eta}{1-x}\(\eta F_{0020}-(1-\eta)F_{0011}\)-4(1-\eta)F_{(-1)021},
\label{F0021lb2}
\end{align}
where $\mathbf{B}=\vec b^2/4$.

Additionally, we have met three integrals that could not be reduced to a combination of known results:  $F_{1110}[(\vec {kb})^2]$, $F_{1110}[(\vec {kb})(\vec {lb})]$, $F_{1110}[(\vec {lb})^2]$. For these integrals we have derived the expressions in the  Schwinger parameterization, and evaluated them in $\epsilon$-expansion up to the finite term following the strategy described in the book \cite{Smirnov:2004ym}.

\section{Logarithm terms of matching coefficient for lpTMDPDF}
\label{app:B}

In this appendix the logarithmic part of the matching coefficients for lpTMDPDFs are collected. Note that these coefficients are not original, in the sense that they can be predicted from the NLO matching derived in \cite{Echevarria:2015uaa,Gutierrez-Reyes:2017glx} via evolution equations as it is described in sec.~\ref{sec:logs}. In our calculation we have derived these expressions directly, as  part of the checks.

Recalling that the perturbative expansion of the coefficient function in eq.~(\ref{OPE_lpG}) is
\begin{eqnarray}
\delta^L\!C_{g\ot f}(x,\vec b;\mu,\zeta,\mu)=\sum_{n=1}^\infty a_s^n \delta^L\!C^{[n]}_{g\ot f}(x,\vec b;\mu,\zeta,\mu),
\end{eqnarray}
with $a_s=g^2/(4\pi)^2$ and  solving the system of eq.~(\ref{RGE4},~\ref{RGE5}) we obtain
\begin{eqnarray}
\delta^L\!C^{[1]}_{g\ot f}(x,\vec b;\mu,\zeta,\mu)&=&\delta^L\!C^{(1,0,0)}_{g\ot f}(x),
\\
\delta^L\!C^{[2]}_{g\ot f}(x,\vec b;\mu,\zeta,\mu)&=&\(-\frac{1}{2}\mathbf{L}_\mu^2+\mathbf{L}_\mu \mathbf{l}_\zeta\)\delta^L\!C^{(2,1,1)}_{g\ot f}(x)+\mathbf{L}_\mu 
\delta^L\!C^{(2,1,0)}_{g\ot f}(x)+\delta^L\!C^{(2,0,0)}_{g\ot f}(x),
\end{eqnarray}
where
\begin{eqnarray}
\mathbf{L}_\mu=\ln\(\frac{\mu^2\vec b^2}{4e^{-\gamma_E}}\),\qquad \mathbf{l}_\zeta=\ln\(\frac{\mu^2}{\zeta}\).
\end{eqnarray}
Using expression for the NLO coefficients (\ref{res:NLO},\ref{res:NLO1}) and the LO DGLAP kernels \cite{Altarelli:1977zs} and expressions for anomalous dimensions (see e.g.\cite{Echevarria:2016scs}) we obtain
\begin{eqnarray}
\delta^L\!C^{(2,1,1)}_{g\ot g}(x)&=&-8C^2_A\frac{1-x}{x},
\\
\delta^L C_{g\ot g}^{(2;1,0)}(x)&=&-16 C_A^2\Big\{\frac{1+x}{x}\ln x+\frac{1-x}{x}\[\frac{x}{6}(2-x)+\frac{15}{4}-\ln(1-x)\]\Big\}
\\\nn && 
+16 C_F T_r N_f \[\frac{1}{3}\frac{1-x}{x}\(2+(2-x)x\)+\ln x\]
 +\frac{16}{3}C_A T_r N_f\frac{1-x}{x},
\end{eqnarray}
\begin{eqnarray}
\delta^L\!C^{(2,1,1)}_{g\ot q}(x)&=&-8C_FC_A\frac{1-x}{x},
\\
\delta^L C_{g\ot q}^{(2;1,0)}(x)&=&-4C_F C_A\[\frac{1-x}{x}\(\frac{43}{3}+x\)+4\frac{1+x}{x}\ln x\]
\\\nn &&
+4C_F^2 \[\frac{1-x}{x}(x+4\ln (1-x))+2\ln x\]
+\frac{32}{3} C_F T_r N_f\frac{1-x}{x},
\end{eqnarray}
where $C_A=N_c$, $C_F=(N_c^2-1)/2N_c$ are Casimir eigenvalues of adjoint and fundamental representation for $SU(N_c)$-gauge group, $T_r=1/2$ is the normalization of Gell-Mann matrices, and $N_f$ is the number of quark flavors.

\bibliographystyle{JHEP}  
\bibliography{TMDref}

\end{document}